\documentstyle[12pt]{article}
\def\lsim{\mathrel{\rlap{\lower4pt\hbox{\hskip1pt$\sim$}}
    \raise1pt\hbox{$<$}}}     
\def\gsim{\mathrel{\rlap{\lower4pt\hbox{\hskip1pt$\sim$}}
    \raise1pt\hbox{$>$}}}     
\begin{document}

\begin{center}
{\Large \bf Five-Dimensional QED, Muon Pair Production and Correction to the Coulomb Potential  }
\vskip .3cm
\normalsize
{  J. R. S. Nascimento\footnote{jroberto@fisica.ufpb.br.},  C. A de S. Pires\footnote{cpires@fisica.ufpb.br.}  and R. F. Ribeiro\footnote{rfreire@fisica.ufpb.br.} }
\vskip .3cm
\it Departamento de F\'{\i}sica, Universidade Federal da
Para\'{\i}ba, Caixa Postal 5008, 58051-970, Jo\~ao Pessoa - PB,
Brazil.
\vskip .3cm
\end{center}
\begin{abstract}
 We consider QED in  five dimensions in a configuration where matter is localized on a 3-brane while  foton  propagates in the bulk. The idea is to investigate the effects of the
 Kaluza-Klein modes of the photon in the relativistic regime, but in low energy, and in the nonrelativistic regime. In the relativistic regime, we calculate the cross section for the reaction $e^+ + e^- \rightarrow \mu^+ + \mu^-$. We  compare our theoretical result with a precise measurement of this cross section at $\sqrt{s}=57.77$ GeV. As result, we extract a lower bound on the size of the extra dimension. In the nonrelativistic regime, we  derive the contribution for the Coulomb potential due to the whole tower of the Kaluza-Klein excited modes of the photon. We use the modified  potential to calculate the Rutherford scattering differential cross section. 

\end{abstract}

\hspace{ 3mm} PACS numbers: 11.10.Kk; 13.66.De
\section{Introduction}

In the Arkani-Hamed-Dimopoulos-Dvali (ADD) original idea of large extra dimensions\cite{add} the standard model(SM) of particle physics was deliberately restricted to a 3-brane while  gravity  propagates in the bulk. The intention behind such arrangment was to solve the hierarchy problem by bringing down the fundamental Planck scale to values around the electroweak scale. However in braneworld scenarios, the lowering of the fundamental Planck scale is compensated by an increasing in the size of the extra dimensions. It is the smallness of the fundamental Planck scale and the largeness of the extra dimensions that has been intensively explored phenomenologicaly over the past five years. In what concern collider phenomenology, basically people has investigated the effect of graviton in standard process through reactions involving  virtual graviton exchange or process with graviton emission\cite{addphen}. 
  
 The largeness of the extra dimensions change considerably if we allow standard field propagates in the bulk. In this context many possibilities were already investigated with different models of particle fields, as for example the standard model\cite{pomarol,carone,rizzo,pilaftsis,universal}, left-right-model\cite{LR}, 3-3-1 model\cite{331}, etc. Focusing only on the standard model(SM), we can have the case where the standard group $SU(2)_L$ is left in the brane, while the group $U(1)_Y$ is put in the bulk\cite{carone}. Other possibility is to put $SU(2)_L$ in the bulk while keeping $U(1)_Y$ in the brane. This last possibility allows many variations. For example, in it we can maintain matter and Higgs fields in the brane or in the bulk. We can have also the whole standard group $SU(2)_L \otimes U(1)_Y$  embeded in the bulk. In this case we still can have variations like keeping part of the particle content in the bulk and  part in the brane, or  all the particle content in the bulk. Even in this case we can have other variations, as for example in models where fermions are localized at specific point along the compact dimension\cite{localized}. In summary, extra dimensions is an extraordinary arena for model building.
 
  The collider phenomenology of all these scenarios  has been intensively studied\cite{review} and, in general,  what has been investigated is the impact of any standard field in the bulk upon the size of the extra dimensions. The result is that with any standard field in the bulk the size of the extra dimensions is pushed down  to value around $R\approx$ TeV$^{-1}$, much smaller than in the original ADD scenario where $R\approx 10^3$ eV$^{-1}$ \footnote{Recently there has appeared an interesting scenario called universal extra dimensions where all the particles of the standard model are put in the bulk\cite{universal}. The peculiarity of this scenario is that  Kaluza-Klein(KK) number is conserved which means that KK modes  are only produced  in pairs. This scenario allows $R$ around hundreds of GeV$^{-1}$.}

In this work we investigate the impact on $R$ due  only to photon in the bulk in a
regime of low energy. To do this we consider a modification of QED called five-dimensional QED(5D QED) where matter field is restricted to live in the 3-brane, while photon is promoted to live in the bulk. The framework of this 5D QED was already set up in Refs. \cite{pilaftsis,santamaria}. In the 5D QED we calculate the contribution of the whole tower of KK modes of the photon to the cross section of the muon pair production reaction. We compare our result  with experimental data at $\sqrt{s}=57.77$ GeV. After we investigate the non-relativistic regime where we analyze the effect of the whole tower of KK modes of the photon on the Coulomb potential. 
 
We organize this paper in the following way. In Sec. \ref{sec1} we present the model. In Sec. \ref{sec2} we calculate the muon pair production cross section and compare our results with experiments. In Sec. \ref{sec3} we obtain the modified Coulomb potential and  with it we calculate the differential cross section for the Rutherford scattering. In Sec. \ref{sec4} we summarize our results and present our conclusions.

\section{5D QED Framework}
\label{sec1}

The 5D QED that we treat here has the following Lagrangian\cite{pilaftsis,santamaria}
\begin{eqnarray}
{\cal L}(x,y)=\bar \Psi(x)(i\gamma^\mu D_\mu  -m)\Psi(x) \delta(y)-\frac{1}{4}F_{MN}(x,y)F^{MN}(x,y),
\label{lag5D}
\end{eqnarray}
where 
\begin{eqnarray}
	F_{MN}(x,y)=\partial_M A_N(x,y) -\partial_N A_M(x,y),
	\label{kinterm}
\end{eqnarray}
and
\begin{eqnarray}
	D_\mu = \partial_\mu -ie_5 A_\mu(x,y),
	\label{covder}
\end{eqnarray}
stand for  the 5-dimensional field strength tensor and the covariant derivative respectively.  The delta function above serves to fix matter field $\Psi$ in the 3-brane. Here we adopt the following notation $M,N,...=0,1,2,3,5$ and $y=x^5$.

First thing to do is to compactify  the  field that inhabit the extra dimension. Here it is favorable to compactify the photon in a $S^1/Z_2$ orbifold with the  fields obeying\cite{pilaftsis}
\begin{eqnarray}
&&	A_M(x,y) = A_M(x,y+2\pi R), \nonumber \\
&&A_\mu(x,y)=A_\mu(x,-y), \nonumber \\
&&A_5(x,y)=-A_5(x,-y).
\label{conditions1}
\end{eqnarray}
With these suppositions the fields $A^\mu(x,y)$ and $A_5(x,y)$ get expanded in Fourier modes in the following way\cite{pilaftsis,santamaria}
\begin{eqnarray}
&&A^\mu(x,y)=\frac{1}{\sqrt{2\pi R}}A^\mu_{(0)}(x)+\sum_{n=1}^{\infty}\frac{1}{\sqrt{\pi R}}A^\mu_{(n)}(x)\cos(\frac{ny}{R}),\nonumber \\
&&A^5(x,y)=\sum_{n=1}^{\infty}\frac{1}{\sqrt{\pi R}}A^5_{(n)}(x)\sin(\frac{ny}{R}).
\label{fieldexpansion}
\end{eqnarray}

The fields $A^\mu_{(n)}(x)$ and $A^5_{(n)}(x)$ are  called the KK excited modes of the original field $A^M(x,y)$. The common procedure at this point is the substitution of  these expansions in (\ref{fieldexpansion})  in the Lagrangian in (\ref{lag5D}), integrate out the fifth dimension and fix apropriately the gauges\cite{gauge}. After all these steps we are left with the following 4-dimensional Lagrangian
\begin{eqnarray}
{\cal L}_4	&=&\bar \Psi(x) (i\partial\!\!\!/ -m)\Psi(x) +e\bar{\Psi}(x)\gamma_\mu \Psi(x) A^\mu_{(0)}(x) -\frac{1}{4}F_{(0)\mu \nu}F_{(0)}^{\mu \nu}  +\nonumber \\
&&+\sqrt{2}e\sum_{n=1}^{\infty}\bar \Psi \gamma_\mu \Psi A^\mu_{(n)}-\frac{1}{4}\sum_{n=1}^{\infty}F_{(n)\mu \nu}F_{(n)}^{\mu \nu}+\sum_{n=1}^{\infty}(\frac{n}{R})^2A^\mu_{(n)}A_{(n)\mu}.
\label{compaclag}
\end{eqnarray}

We now discuss  some generic features of the couplings and fields that appears in the 4-dimensional Lagrangian above. First thing to note is that it is implicit in the Lagrangian above the redefinition $e=\frac{e_5}{2\pi R}$ of the strong coupling $e_5$ by the standard electromagnetic coupling $e$. See also that the field $A^0_\mu$ remains massless and that the first three terms in the lagrangian above recover the usual QED Lagrangian. This shows that $A^0_\mu$ is the photon. Note also that all KK modes $A^n_\mu(x)$  are  vectorial bosons, while that all $A^n_5(x)$ are scalar bosons. Perceive that in the  Lagrangian in (\ref{lag5D}) the original vectorial boson $A_M(x,y)$ is massless. It is only after the compactification of the fifth dimension that all $A^n_\mu(x)$ (with $n\geq1$) gain mass, while  all $A^n_5(x)$ (with $n\geq1$) remain massless and disappear from the 4-dimensional Lagrangian in (\ref{compaclag}). This is achieved through a carrefull fixing of the gauge\cite{pilaftsis}. It seems that the fields $A^n_5(x)$ (with $n\geq1$) play the role of  Goldstone bosons in a kind of geometric version of the Higgs mechanism.  At the end, the  picture appears as if the composed symmetry $G=\displaystyle \prod_{n=1}^{\infty} U^n(1)$  is broken to the ordinary $U(1)_{em}$ symmetry.

\section{ Muon pair production }
\label{sec2}
In this section we calculate the cross section for muon pair production in the framework of the 5D QED. Our main motivation is to confront the theory with experiments. What justify the choice of this process is its simplicity once it presents only contributions through $s$ channel. 

The muon pair production is the reaction
\begin{eqnarray}
e^+(p_1)+ e^-(p_2) \rightarrow \mu^+(k_1 ) +\mu^-(k_2) 	
\label{process}
\end{eqnarray}
Here we use the  Feynman rules of Ref. \cite{pilaftsis}.  According to those rules,  we get the following scattering amplitude for  muon pair production in unitary gauge
\begin{eqnarray}	
i{\cal M}=ie^2\bar{u}(k_1)\gamma^\mu v(k_2)\left( \frac{ g_{\mu \nu} }{ q^2 }+2\sum_{n=1}^{\infty}\frac{ g_{\mu \nu} -\frac{q_\mu q_\nu}{m^2_n} }{q^2 -m^2_n  }  \right)\bar{v}(p_2)\gamma^\nu u(p_1),
\label{amplitude}
\end{eqnarray}
where $m_n=n/R$ is the mass of the n-th KK mode. The two terms in the parenthesis above are due to the interchange, via $s$ channel, of the photon and of the whole tower of KK modes, respectively. See that we have to handle infinite contributions of KK modes. What we have to do is to develop the sum that appears in the amplitude above. It can be expanded to give
\begin{eqnarray}
2\sum_{n=1}^{\infty}\frac{ g_{\mu \nu} -\frac{q_\mu q_\nu}{m^2_n} }{q^2 -m^2_n  }&=&g_{\mu \nu} \left( -\frac{1}{q^2}+\frac{\pi}{qM_c}\cot(\frac{\pi q}{M_c})\right)- \nonumber \\
&&\frac{q_\mu q_\nu}{q^2}\left(-\frac{1}{q^2}+\frac{1}{3}(\frac{\pi}{M_c})^2 +\frac{\pi}{qM_c}\cot(\frac{\pi q}{M_c})\right),
\label{expanding}
\end{eqnarray}
where $M_c=1/R$. We collect the terms in leading order in $M_c$ to get
\begin{eqnarray}
2\sum_{n=1}^{\infty}\frac{ g_{\mu \nu} -\frac{q_\mu q_\nu}{m^2_n} }{q^2 -m^2_n  }=\frac{1}{3}\left( g_{\mu \nu} -2\frac{q_\mu q_\nu}{q^2}\right)(\frac{\pi}{M_c})^2.
\label{expansion}	
\end{eqnarray}

Even though we are restricted to low energy regime, it is still high enough  to allow to use the limit $\sqrt{s}> m_\mu$. In view of this the center of mass frame get characterized by
\begin{eqnarray}
&&	p_1=\frac{\sqrt{s}}{2}(1,0,01)\,\,\,, p_2=\frac{\sqrt{s}}{2}(1,0,,0,-1)\,\,\,,\nonumber \\
&&k_1=\frac{\sqrt{s}}{2}(1,\sin\theta,0,\cos\theta)\,\,\,,k_2=\frac{\sqrt{s}}{2}(1,-\sin\theta,0,-\cos\theta).
\label{momentum2}
\end{eqnarray}
 
After these considerations,  we are now able to obtain the unpolarized squared amplitude. We substitute (\ref{momentum2})  and (\ref{expansion}) in (\ref{amplitude})  to obtain
\begin{eqnarray}
|{\cal M}|^2 = e^4\left(1+ \frac{2s}{3}(\frac{\pi}{M_c})^2\right)(1+\cos^2\theta).
\label{amplsquared}
\end{eqnarray}
This result leads to 
\begin{eqnarray}
\frac{d\sigma}{d\Omega}=\frac{\alpha^2}{4s}\left(1+\frac{2\pi^2 s}{3M^2_c}\right)\left(    1+\cos^2\theta \right),
\label{difcrossect}	
\end{eqnarray}
for the differential cross section, and
\begin{eqnarray}
\sigma_{tot}=\frac{4\pi \alpha^2}{3s}\left(1+\frac{2\pi^2 s}{3M^2_c}\right) .
\label{totalcrosct}	
\end{eqnarray}
for the total cross section. The first term in the total cross section is the one predicted by ordinary QED. The second term is due to the contributions of infinite KK modes. See that we recover the ordinary QED prediction when $R$  goes to zero.  

There is a model independent precise measurement of this cross section at $\sqrt{s}=57.77$ GeV which obtained for the total muon pair production cross section the value\cite{measurement}
\begin{eqnarray}
	\sigma_{exp}=30.05 \pm 0.59 \mbox{pb}.
	\label{experimet}
\end{eqnarray}

We have now all the ingredients that permit us to check how relatively sensitive is the 5D QED for large extra dimension. For this we confront (\ref{experimet})  with (\ref{totalcrosct}). The result is the lower bound
\begin{eqnarray}
	M_c \geq 689 \mbox{GeV\,\,\,\,95\% CL}.
	\label{valuetoR}
\end{eqnarray}
This bound is not so far from other bounds obtained in scenarios with the whole or part of the  SM field content put in the bulk \cite{pomarol,carone,rizzo,pilaftsis}. Thus we can conclude
that, despite 5D QED is the simplest scenario which involve standard gauge field in the bulk, it is still as sensitive for large extra dimension as other more complex scenarios\cite{pomarol,carone,rizzo,pilaftsis}.

\section{ Modified Coulomb Potential }
\label{sec3}

The issue we attack in this section is  what correction  the whole tower of KK modes induces in the Coulomb potential? This requires that we  go to the non-relativistic regime. 

In this section we proceed as in Ref. \cite{peskin}. Thus the Coulomb potential will be the Fourier transform of the non-relativistic limit of the temporal parte of the scattering amplitude, ${\cal M}^0$, of a general reaction
\begin{eqnarray}
	V(x)= \int\frac{d^3q}{(2\pi)^3}{\cal M}^0 e^{i\vec{q}.\vec{x}},
	\label{ftransf}
\end{eqnarray}

Let us take the following general reaction $l(p) + l(k) \rightarrow l^{\prime}(p^{\prime}) + l^{\prime}(k^{\prime})$ such that the fermions be distinguishable. Its scattering amplitude is
\begin{eqnarray}	
i{\cal M}=ie^2\bar{u}(p^{\prime})\gamma^\mu u(p)\left( \frac{ g_{\mu \nu} }{ q^2 }+2\sum_{n=1}^{\infty}\frac{ g_{\mu \nu} -\frac{q_\mu q_\nu}{m^2_n} }{q^2 -m^2_n  }  \right)\bar{u}(k^{\prime})\gamma^\nu u(k),
\label{2amplitude}
\end{eqnarray}
where $q=k^{\prime}-k$. The two terms in parenthesis are due to $t$ channel interchange of photon and the whole tower of KK modes, respectively. 

To evaluate any amplitude in the nonrelativistic limit, we have to keep terms until to lowest order in the 3-momenta\cite{peskin}. In view of this the common approximation for the momenta is 
\begin{eqnarray}
	p=(m,\vec{p}),\,\,\, p^{\prime}=(m,\vec{p^{\prime}}),\,\,\,k=(m,\vec{k}),\,\,\,k^{\prime}=(m,\vec{k^{\prime}}).
	\label{momenta}
\end{eqnarray}
This implies that $q^2=-\vec{q}^2$ and therefore the temporal part of the amplitude above take the form
\begin{eqnarray}
	i{\cal M}^0=-ie^2\bar{u}(k_1)\gamma^0 u(k_2)\left( \frac{ 1 }{ \vec{q}^2 }+2\sum_{n=1}^{\infty}\frac{ 1 }{\vec{q}^2 +m^2_n  }  \right)\bar{u}(p_2)\gamma^0 u(p_1).
\label{0amplitude}
\end{eqnarray}
Before taking the Fourier transform, let us evaluate the sum in (\ref{0amplitude}). It takes the following simple expression
\begin{eqnarray}
\sum_{n=1}^{\infty}\frac{ 1 }{\vec{q}^2 +m^2_n  }= \frac{1}{6}(\frac{\pi}{M_c})^2.
\label{sum2}	
\end{eqnarray}

We now substitute (\ref{sum2}) in (\ref{0amplitude}), and we take the non-relativistic limit and  its  Fourier transform as in (\ref{ftransf}). Doing this we obtain the following expression for the Coulomb potential in leading order in $M_c$
\begin{eqnarray}
	V(r)=\frac{\alpha}{r}+\frac{4}{3}\frac{ \alpha \pi^3 }{M^2_c} \delta(r),
	\label{potential}
\end{eqnarray}
where $\delta(r)=\delta(x)\delta(y)\delta(z)$. Therefore the effect of the whole tower of KK modes  is a correction of the classical Coulomb potential in form of a  delta function of the position. The delta function can be easily understood if we perceive that it is implicit in the approximation in (\ref{sum2}) that the KK modes are very heavy in agreement with the lower bound in (\ref{valuetoR}). As consequence each matter-KK mode interaction is in fact an interaction of contact. The delta function is a consequence of infinite interactions of contact. This delta means a very strong interaction in the origin which translates in bound state when the potential is attractive, or in an infinite barrier in case of repulsive potential. This delta potential is not a novelty in physics. It arises naturaly when we take the nonrelativistic limit in interaction terms like $\lambda \phi^4$\cite{jackiw}. 

As an application of this modified Coulomb potential, let us calculate the correction for the Rutherford scattering differential cross section.

To do this we first have to obtain the partial wave amplitude in first order in Born approximation. It is the Fourier transform of the Coulomb potential
\begin{eqnarray}
	f(\theta)=-\frac{\mu}{2\pi}\int d^3r V(r)e^{i\vec{q}.\vec{r}}.
	\label{pwampl}
\end{eqnarray}
 Substituting (\ref{potential})  in (\ref{pwampl}) and integrating out we get
 \begin{eqnarray}
f(\theta)=\frac{2 \mu \alpha}{q^2}+\frac{2}{3}\frac{\mu \alpha \pi^3}{M^2_c} 	
\label{partialwave}
\end{eqnarray}

Let us suppose that an incident beam coming with momentum $\vec{k}$ being scattered by a target and getting momentum $\vec{k^{\prime}}$. The target recoil is $\vec{q}=\vec{k^{\prime}}-\vec{k}$. Expressing  this in terms of the energy $E$ and the scattering angle $\theta$, we have
\begin{eqnarray}
	|\vec{q}|^2=8\mu E \sin^2(\theta/2),
	\label{momentum}
\end{eqnarray}
 where $\mu$ is a reduced mass and $\theta$ is the angle between $\vec{k}$ and $\vec{k^{\prime}}$. With the partial-wave amplitude in (\ref{partialwave}) and the expression for the momentum transferred in (\ref{momentum}),  we get the following expression in leading order in $M_c$ for the Rutherford scattering of a incident beam of charge $Z_1 e$ with a target of charge $Z_2e$,
\begin{eqnarray}
	\frac{d\sigma}{d\Omega}=|f(\theta)|^2=\frac{Z_1 Z_2\alpha^2}{16 E^2 \sin^4(\theta/2)}+\frac{4Z_1 Z_2 \alpha^2 \mu \pi^3 }{3EM^2_c \sin^2(\theta/2)}.
	\label{rutherford}
\end{eqnarray}

 As we can read from the expression above, correction for the classical Rutherford scattering appears in the same order in $M_c$ as it appears  in corrections for obervables in high energy. 

\section{Summary and Conclusions}
\label{sec4}

In the first part of this work we explored phenomenologicaly the so-called 5D QED. We calculated the contributions of this theory for the reaction $e^+ + e^- \rightarrow \mu^+ +\mu^-$ and compared it with a precise measurement of such reaction at $\sqrt{s}=57.77$ GeV. The result of such comparison was a lower bound $M_c\geq 689$GeV  on the size of the fifth dimension. This bound is not better than other bounds with part or the whole SM fields in the bulk, however we have to have in mind that 5D QED is an effective theory and that in the scale of energy we considered here contribution from $Z^0$ is small but could be decisive to improve the bound on $M_c$. For our proposal of only investigating bound on extra dimension with only the photon  propagating in the bulk, we think that  the bound we found in (\ref{valuetoR}) is revealing enough of the sensibility of the extra dimensions for the case of standard gauge field propagating in the bulk.

In what followed,  we  examined  the effect of extra dimension in the nonrelativistic regime. We found that  the whole tower of KK modes modifies the Coulomb potential by a term dependent on a  delta function of the position. We then applied the modified Coulomb potential to calculate the classical Rutherford scattering differential cross section. We obtained a deviation from the classical Rutherford scattering differential cross section  in the same order in $M_c$ as it appears in  observables in high energy. 

To finalize, phenomenology of extra dimensions has focused attention mainly in high energy physics and neglected the consequences of large extra dimensions at the low energy physics. We followed the second line and investigated some  consequences of large extra dimensions at low energy scale.  We think that the results obtained in this work indicate that lower energy physics can be viable for constraining large extra dimensions.


{\bf Acknowledgments.} We thank S. Lietti for valuable discussions and  D. Bazeia for reading  the manuscript.   J.N and C.P thank CNPq for partial support. We also thank PROCAD  for partial support.

\end{document}